\documentstyle[11pt,aaspp4]{article}
\slugcomment{}

\lefthead{}
\righthead{}

\begin{document}

\title{The Evolution of the Global Star Formation History as Measured
from the Hubble Deep Field\footnote{
Based on observations made with the NASA/ESA
Hubble Space Telescope, obtained from the data archive at the
Space Telescope Science Institute, which is operated by the
Association of Universities for Research in Astronomy, Inc., under
cooperative agreement with the National Science Foundation}
}

\author{A.J. Connolly, A.S. Szalay, Mark Dickinson\altaffilmark{2,3},
M.U. SubbaRao and R.J. Brunner} 

\affil{Department of Physics and Astronomy, The Johns Hopkins
University, Baltimore, MD 21218\\ Electronic mail:
ajc@skysrv.pha.jhu.edu, szalay@skysrv.pha.jhu.edu, med@stsci.edu,
subbarao@skysrv.pha.jhu.edu, rbrunner@skysrv.pha.jhu.edu}

\altaffiltext{2}{Visiting observer at the Kitt Peak National
Observatory, National Optical Astronomy Observatories, which is
operated by the Association of Universities for Research in Astronomy,
Inc. (AURA) under cooperative agreement with the National Science
Foundation.}
\altaffiltext{3}{Space Telescope Science Institute}

\begin{abstract}

The Hubble Deep Field (HDF) is the deepest set of multicolor optical
photometric observations ever undertaken, and offers a valuable data 
set with which to study galaxy evolution. Combining the optical 
WFPC2 data with ground--based near--infrared photometry, we derive 
photometrically estimated redshifts for HDF galaxies with $J<23.5$.  
We demonstrate that incorporating the near--infrared data reduces 
the uncertainty in the estimated redshifts by approximately 40\% and 
is required to remove systematic uncertainties within the redshift 
range $1<z<2$. Utilizing these photometric redshifts, we determine 
the evolution of the comoving ultraviolet (2800\AA) luminosity density 
(presumed to be proportional to the global star formation rate) from 
a redshift of $z=0.5$ to $z=2$. We find that the global star formation 
rate increases rapidly with redshift, rising by a factor of 12 from 
a redshift of zero to a peak at $z \approx 1.5$.  For redshifts beyond 
1.5, it decreases monotonically.  Our measures of the star formation 
rate are consistent with those found by Lilly et al.\ (1996) from the 
CFRS at $z < 1$, and by Madau et al.\ (1996) from Lyman break galaxies 
at $z > 2$, and bridge the redshift gap between those two samples.  
The overall star formation or metal enrichment rate history is consistent 
with the predictions of Pei and Fall (1995) based on the evolving HI 
content of Lyman-$\alpha$ QSO absorption line systems.

\end{abstract}


\section{Introduction}

The multicolor photometric survey of the Hubble Deep Field (HDF;
Williams et al.\ 1996) provides a unique opportunity for studying the
evolution of the global star formation history of the Universe.
Quantifying this relation has significant consequences for the
evolution of structure formation, supernova rates and the metal
enrichment of the intergalactic medium as a function of epoch.  For
redshifts less than one, the Canada France Redshift Survey (CFRS;
Lilly et al.\ 1996) has shown that the comoving ultraviolet (2800\AA)
luminosity density of galaxies increases rapidly as a function of
redshift, rising by a factor of $\sim$15 from $z=0$ to 1.  Because the
UV luminosity of star--forming galaxies is, broadly speaking,
proportional to their star formation rates, this change in luminosity
density has been inferred to trace a global decline in the cosmic star
formation rate since $z=1$.  Using the 1500\AA\ luminosity of galaxies
in the HDF selected by the presence of a Lyman continuum break, Madau
et al.\ (1996) find that the inferred star formation rate at $z > 2$
is lower than the value measured by the CFRS at $z \approx 1$, and
falls steadily toward higher redshifts.  The implication of these two
surveys is that the comoving luminosity density peaks somewhere within
the redshift range $1<z<2$.  To date, however, there have been no
systematic surveys of this redshift range, and consequently the
evolution of the luminosity density for $1<z<2$ remains a matter of
conjecture.  Based on the evolution of the global HI content of the
universe as measured from Lyman~$\alpha$ QSO absorption line systems,
Pei \& Fall (1995) have predicted that the cosmic star formation rate
should peak somewhere in this redshift range.  It therefore seems
worthwhile to undertake a study of the global star formation rate
within this redshift regime. Throughout this letter we assume an
$\Omega=1, q_0 = 0.5$ cosmology.

\section{Near Infrared Observations of the HDF}

To complement the optical multicolor data of the HDF, followup
near-infrared observations were undertaken with the KPNO 4m
telescope.   Full details of the observations and data reduction
will be described in Dickinson et al.\ (1997);  we briefly summarize 
the salient features here.  

At the f/15 focus of the Mayall 4m, the IRIM 256x256 Rockwell NICMOS-3 
HgCdTe infrared array covers a 2.56 arcmin field-of-view with a 0.6 arcsecond
pixel scale. The field-of-view of IRIM is, therefore, well matched to
the WFPC2 images. Over the course of 10 days, a total of 11.0, 11.3 and
22.9 hours of data were collected for the HDF in the $J$, $H$ and $K_s$ 
passbands. The formal magnitude limits for a two arcsecond aperture 
and a signal-to-noise ratio of five are 23.45, 22.29, and 21.92 in
the $J$, $H$ and $K_s$ filters respectively.

All images were combined using a ``drizzling'' method similar to that
applied to the optical HDF data (Fruchter \& Hook, 1997).  This resulted 
in a plate scale of 0.1594 arcsec pixel$^{-1}$. The near-infrared images 
were registered and transformed to the same geometry as the optical WFPC2 
frames (after binning down the optical images to the pixel scale of the near
infrared data). 

\section{Optical and Near--Infrared Photometric Catalogs}

For the purposes of this paper, we wish to select galaxies from the
HDF in a fashion similar to that used for the CFRS survey which has 
been used to estimate the redshift evolution of global star formation
rates at $z < 1$.   The CFRS galaxy sample was derived from $I$--band 
photometry so as to select objects on the basis of their rest--frame 
optical ($\lambda_0 > 4000$\AA) emission out to $z \approx 1$.  
In order to broadly reproduce this strategy at higher redshift, 
we have chosen our HDF galaxy sample in the $J$--band (i.e. at 
$\lambda_{\rm obs} \approx 12500$\AA).  This ensures that in the 
$1 < z < 2$ redshift range of interest here, galaxies are selected 
at similar rest--frame wavelengths as the $I$--band CFRS probes for 
$0.2 < z < 1$. The $J$--band is approximately equivalent to a 
redshifted rest--frame $B$--band at $z = 1.9$.

Photometric catalogs for the optical and near-infrared data were
constructed using a modified version of the SExtractor image analysis
program (Bertin \& Arnouts 1996).  Object detection was carried out
using a 1 arcsec FWHM Gaussian detection kernel on the $J$ image.
Matched aperture photometry was performed on each of the optical and
near infrared images. A 4 arcsec diameter aperture was placed on each
object detected in the $J$--band data. To reduce the effect of flux
from nearby objects contaminating these aperture magnitudes,
SExtractor was modified to mask out nearby sources (Brunner 1997).

The resulting optical and near--infrared catalog contains a total of
219 galaxies to a magnitude limit of $J<23.5$.  At $z=2$ for our
adopted cosmology, this $J$--band apparent magnitude limit corresponds
to a rest--frame absolute magnitude of $B = -20.4$, or approximately
0.6 magnitudes fainter than present--day $L^\ast$ (i.e. assuming no
luminosity evolution).  Of the cataloged sources, 194 and 204 were
detected independently in the $H$ and $K_s$ passbands.  At these
signal--to--noise levels, all objects in our infrared catalog have
optical counterparts in the WFPC2 images.  Of these, 73 have measured
spectroscopic redshifts (Cohen et al.\ 1996, Steidel et al.\ 1996,
Lowenthal et al.\ 1996).  Their redshift distribution is shown in
Figure 1a.

\section{An Empirical Approach to Photometric Redshifts}

A number of attempts have been made to estimate the redshifts and
spectral types of galaxies from broadband photometry using
ground--based data (Koo 1986, Connolly et al.\ 1995). The success of
these studies led to the application of photometric-redshifts to
optical data for the HDF (Gwyn and Hartwick 1996, Mobasher et al.\
1996, Lanzetta et al.\ 1996, Sawicki et al.\ 1997). In the above
references, a comparison between the spectroscopic and photometric
redshifts has been shown to result in a dispersion of $\sigma_z \sim
0.15$ in the redshift ranges $0<z<1$ and $2.2<z<3.5$.  However, there
are significantly larger errors in the $1<z<2$ regime.

The increase in the dispersion for $z>1$ is due to the HDF passbands
sampling the optical and near--UV rest frame wavelengths. The
dominant feature that provides the photometric redshift is the
transition of the break in the galaxy continuum around 4000 \AA\
through the F450W and F606W filters (Connolly et al.\ 1995). At $z>1$
this break moves out of the optical spectral region and into the
near-infrared. For star--forming galaxies, the ultraviolet continuum
from Lyman~$\alpha$ (1215\AA) to $\sim$3000\AA\ is relatively devoid
of strong features, and consequently there is little information in
optical photometry from which to estimate redshifts at $1 < z < 2$. At
redshifts greater than 2.2, the 912\AA\ Lyman limit enters the U band,
and redshifts can again be estimated in an analogous manner to the
4000\AA\ discontinuity (Steidel et al.\ 1996, Madau et al.\ 1996).

Therefore, using only the HDF optical data, the redshift range from
$1<z<2$ is poorly constrained. With the addition of deep $J$--band
data alone, we can detect the transition of the 4000\AA\ discontinuity
out to $z=2.1$. To demonstrate this, in Figure 2 we fit a quadratic
relation between the HDF optical magnitudes (only) and the
spectroscopic redshifts (for galaxies with $z<1.5$). The dispersion
about this relation is $\sigma_z = 0.097$. The dispersion increases as
a function of redshift, doubling by a redshift of one. If we
incorporate the near-infrared $J$ band data and refit the relation,
the dispersion decreases by 40\% ($\sigma_z = 0.059$). Similar success
has been reported by Cowie (1996; see also Ellis 1997) using an
independent infrared data set and a different redshift estimation
methodology.

To derive the estimated photometric redshifts out to $z=2$, we fit a
3rd order polynomial in the F300W, F450W, F606W, F814W and $J$
passbands to the observed spectroscopic data (full details of our
technique can be found in Connolly et al.\ 1995 and Brunner et
al. 1997). The correlation between the spectroscopic and photometric
redshifts is given in Figure 2. The dispersion within this relation is
$\sigma_z =0.06$.  Applying this relation to those galaxies without
spectroscopic redshifts, we derive the photometric redshift sample.
In Figure 1b we show a comparison between the observed redshift
distribution for the spectroscopic data and the redshift distribution
determined from the 3rd order fit to the optical and near-infrared
data.

One advantage of our technique is that it does not depend on the
assumption of any particular set of template spectral energy
distributions for the galaxies (cf. Ellis 1997).  The corresponding
limitation to the method is that it requires a calibrating sample of
galaxies with spectroscopic redshifts. In the redshift interval $1 < z
< 2$ there are currently only a handful of galaxies with spectroscopic
redshifts. We, therefore, utilize those galaxies with $z > 2$ to
constrain the high--redshift portion of the photometric redshift
relation; we do not use them in our subsequent analysis of the evolution
of the comoving luminosity density.

\section{Evolution of the Global Star Formation Rate}

The evolution of the comoving UV luminosity density has been
characterized by Lilly et al.\ (1996) for the redshift interval
$0<z<1$, and by Madau et al.\ (1996) for $z>2$.  Lilly et al.\
derived 2800\AA\ luminosity densities for the CFRS galaxy sample,
while Madau et al.\ used 1500\AA\ luminosities for Lyman break
galaxies in the HDF.  Madau et al.\ employ spectral population synthesis 
models to compute conversions from luminosity at these two UV wavelengths 
to star formation rates (SFR) and metal enrichment rates (MER), permitting 
comparison of these quantities for galaxy samples at widely varying redshifts.
The conversion to MER is less sensitive to the assumed form of the 
initial mass function than is the conversion to SFR.  

Using the photometric redshift sample, we investigate the redshift 
dependence of the star formation rate in the range $0.5<z<2$.  This 
lower redshift limit ensures that the HDF optical passbands always sample 
rest frame wavelengths $\lambda > 2000$\AA.  Galaxy luminosities at 
2800\AA\ can therefore be interpolated directly from the observed multicolor 
photometry.  This allows us to compare our results directly with those of 
Lilly et al.\  The upper redshift limit is imposed because at redshift 
$z>2$ the break in the galaxy spectral energy distribution at 4000\AA\ 
moves out of the $J$ band (see \S 4 above).

We construct the comoving luminosity density for three redshift ranges:
$0.5<z<1$, $1<z<1.5$ and $1.5<z<2$. The observed luminosity densities
in these redshift intervals are given in column (2) of Table 1.  Because 
our sample is limited to galaxies with $J < 23.5$, it samples different 
luminosity ranges at each redshift.  Following the procedure of Lilly et al., 
we correct for the incompleteness in the observed luminosity densities
by integrating over a Schechter luminosity function for each redshift bin.  
Column (3) of Table 1 gives the value of $M^\ast$ (at 2800\AA, in AB 
magnitudes) derived from the photometric data for each of the redshift 
intervals.  Columns (4), (5) and (6) give the corrected comoving luminosity 
density assuming a Schechter luminosity function with a slope of 
$\alpha=-1.0$, -1.3 and -1.5 respectively.  For the highest redshift 
bin and the steepest value of $\alpha$, our completeness correction is 
approximately a factor of 2.

With our current data set we cannot reliably determine the evolution of 
the faint end slope of the luminosity function with redshift.  For
comparison with the CFRS sample, we assume a value of $\alpha=-1.3$ for 
the slope of the Schechter function (Lilly et al.\ 1996).  The uncertainties 
in the derived values are estimated by considering the magnitude of the 
incompleteness corrections for a range in $\alpha$ of $-1.0<\alpha<-1.5$.  
For each redshift interval these corrections give an uncertainty of 
approximately 0.15 in the log.  We therefore adopt this value for each 
of the three redshift intervals.

The derived 2800\AA\ luminosity densities were converted to metal enrichment
rates to compare with the predictions of Pei and Fall (1995) and with
the $z > 2$ data of Madau et al.\ (1996) (which, as noted, are based on 
1500\AA\ luminosity densities).   We derive this conversion using the
Bruzual \& Charlot (1996) stellar population synthesis models, assuming
a Salpeter initial mass function, constant star formation rate, solar 
metallicity, and a galaxy age of 0.1--1 Gyr (Dickinson et al.\ 1997, 
Madau et al.\ 1996).  The conversion from L(2800\AA) to metal enrichment 
rate is $2.2\times 10^{-23}$ M$_\odot$ yr$^{-1}$ W$^{-1}$ Hz.  
Note that this differs from the value adopted by Madau et al.\ (1996) 
by a factor of approximately 1.6 due, in part, to changes in the stellar 
synthesis models of Bruzual and Charlot (1996).  The conversion from
1500\AA\ luminosity to MER (which is used for the $z > 2$ data points
for the Lyman break galaxies) remains unchanged from the values used 
by Madau et al.  These changes in the conversion factors go in the sense 
of reducing the MERs or SFRs at $0 < z < 2$ relative to those 
at $z > 2$.

The MER rate as a function of redshift from Gallego et al. (1995; open
triangle), Lilly et al.\ (1996; open circles), Madau (1996; open
squares) and our photometric redshift sample (solid circles) is given
in Figure 3. Superimposed on these plots are the predictions of Pei
and Fall (1995) based on the evolution of the HI content in damped
Lyman-$\alpha$ systems. The solid line represents models with a
comoving gas density, $\Omega_{g\inf} = 4\times$10$^{-3} h^{-1}$, and
the dashed lines with $\Omega_{g\inf} = 2\times$10$^{-3}
h^{-1}$. Models assuming a closed box or outflow model are indicated
by C and O and those assuming an inflow model by I.

\section{Discussion and Conclusions}

The spectroscopic, photometric and Lyman break galaxy samples provide
a remarkably consistent picture of the evolution of the metal
enrichment rate as a function of redshift. Considering these data as a
whole, we find that the comoving luminosity density rises rapidly as
we look back in cosmic time from a redshift of zero to a peak at
approximately $z=1.5$.  It then falls by a factor 2 out to a redshift
of 3.

The amplitude and shape of the observed MER evolution with redshift
are broadly consistent with the models of Pei and Fall (1995).  The models 
with normalizations $\Omega_{g\inf} = 4\times$10$^{-3} h^{-1}$ and
$\Omega_{g\inf} = 2\times$10$^{-3} h^{-1}$ bound the data points.
Given the errors in the observed quantities (and in the conversions
from luminosity density to MER), it is not currently feasible to try to
quantify whether the data best fit a closed, open or infall model.

The interpretation of our results depends on the absence of systematic
errors in the photometric redshift technique, and on understanding the
conversion of ultraviolet luminosity to MER or SFR.  The close
agreement between our photometric and spectroscopic redshifts (Figure
2) encourages us to believe that the redshift estimates are good.  The
relative conversions between MER or SFR and luminosities at 1500\AA\
(for $z > 2$), 2800\AA\ (at $0.2 < z < 2$), and H$\alpha$ ($z = 0$,
from Gallego et al.) depend somewhat on the assumptions of the
population synthesis models.  A potentially more serious effect is
that of dust extinction, which may differ substantially at the various
wavelengths considered here.  The role of extinction in high redshift
galaxies has yet to be comprehensively analyzed from existing data
sets.  While extinction may be expected to have greater impact on the
measurements at 1500\AA, and hence on the $z > 2$ galaxies, one must
also consider the redshift evolution of the dust content of galaxies.
We do not attempt to disentangle this problem here, but call attention
to it as a caution to be kept in mind.

Regardless of the absolute conversion factors, it is important to note
that the comoving MERs which we calculate for the photometric redshift
sample are derived in the same manner as was done for the CFRS, using
luminosity densities at 2800\AA, providing a {\it relatively}
consistent methodology across the very large redshift range $0.2 < z <
2$. It is encouraging that in the overlap region between the
spectroscopic and photometric redshift sample the luminosity densities
are comparable.  Moreover, there is no evidence for a gross mismatch
at the $z=2$ transition between our data set and the Madau et al.\
Lyman break galaxy sample.  We therefore believe that the general form
of the MER evolution over a very broad cosmic timespan is reliably
described by these observations, which, for the first time, span the
peak era in star forming activity in the universe.

\acknowledgments 

We thank Mike Fall, Yichuan Pei and Piero Madau for many useful
discussions on the interpretation of the metal enrichment rate and
comoving luminosity density relation, and Stephane Charlot for
providing the newer population synthesis models.  We particularly
thank the other investigators involved in the planning, observations
and data reduction of the near--infrared imaging data, especially Matt
Bershady, Peter Eisenhardt, Richard Elston, and Adam Stanford. We
acknowledge partial support from NASA grants AR-06394.01-95A and
AR-06337.11-94A (AJC), AR-06337.16-94A (MD), and an LTSA grant (ASZ).

\appendix

\begin{deluxetable}{lccccc}
\tablecaption{Comoving 2800\AA\ Luminosity Density} 
\tablecolumns{6}
\tablewidth{0pc}
\tablehead{
\colhead{Redshift} & \colhead{Observed$^1$} & \multicolumn{4}{c}{LF Corrected$^1$} \\ 
\colhead{} & \colhead{} &  \colhead{$M^\ast$} & \colhead{$\alpha=-1.0$} & \colhead{$\alpha=-1.3$} & \colhead{$\alpha=-1.5$} \\
}
\startdata

0.5 -- 1.0  &   19.30   &  $-20.25$ & 19.43 & 19.52 & 19.65 \\
1.0 -- 1.5  &   19.48   &  $-21.00$ & 19.55 & 19.69 & 19.85 \\
1.5 -- 2.0  &   19.38   &  $-21.75$ & 19.39 & 19.59 & 19.62 \\
\enddata
\tablecomments{(1) log L(2800\AA) in $h_{50}^{-2}$ W Hz$^{-1}$ Mpc$^{-3}$}
\end{deluxetable}

\clearpage
\figcaption{The redshift distribution for those galaxies within the
HDF with $J<23.5$.  Figure 1a shows the distribution for those
galaxies with measured spectroscopic redshifts, while Figure 1b shows
the derived redshift distribution for the photometric redshift
sample. The peak in the n(z), at $z \sim 0.9$, for our
photometric-redshift sample is in reasonable agreement with that
predicted from the gravitational lensing (Ebbels et al.\ 1997).}

\figcaption{A comparison between the spectroscopic and photometric
redshifts for the HDF sample. The inset panels show the effect of
including the near--infrared data into the photometric redshift
relation. The top inset shows the derived relation for the F300W,
F450W, F606W and F814W passbands (for galaxies with $z<1.5$). The
dispersion in the relation is $\sigma_z = 0.097$. The lower inset
shows the dispersion for the F300W, F450W, F606W, F814W and $J$
passbands. The dispersion decreases to $\sigma_z = 0.059$. The
correlation for all galaxies within the HDF ($J<23.5$) is given in the
main figure.}

\figcaption{The metal enrichment rate as a function of redshift as
measured from Gallego et al. (open triangle), Lilly et al. (open
circle), Madau (open square) and the photometric redshift sample
(closed circle). The photometric redshift luminosity densities have
been corrected for incompletion assuming a Schechter luminosity
function with a slope of $\alpha=-1.3$. The solid and dashed lines
represent the predictions of Pei and Fall for the metal enrichment
rate based on the comoving HI density traced by Lyman-$\alpha$ absorption
systems. See the text for details on the individual models.}

%\plotone{fig1.ps}

%\plotone{fig2.ps}

%\plotone{fig3.ps}


\begin{references}

\reference{} Bertin, E., \& Arnouts, S. 1996, A\&A, 117, 393

\reference{} Brunner, R.J., 1997, PhD Thesis, Johns Hopkins University

\reference{} Brunner, R.J., Connolly, A.J., Szalay, A.S. \& Bershady,
M.A., 1997, ApJ in press

\reference{} Bruzual, G. \& Charlot, S., 1996, private communication. 

\reference{} Connolly, A.J.,Csabai, I., Szalay, A.S., Koo, D.C., Kron,
R.C., Munn, J.A., 1995, AJ, 110, 2655

\reference{} Cowie, L.L., 1996, The proceedings of the 37th
Herstmonceux conference on ``HST and the high redshift universe,''
   in press 

% \reference{} Cowie, L.L., Clowe, D., Fulton, E., Cohen, J.G., Hu, E.M.,
% Songaila, A., Hogg, D.W., and Hodapp, K.W., 1997, in preparation.

\reference{} Cohen, J.G., Cowie, L.L., Hogg, D.W., Songaila, A., Blandford, R.,
Hu, E.M. \& Shopbell, P., 1996, ApJ, 471, L5.

\reference{} Dickinson et al.\ 1997, in preparation

\reference {} Ebbels, T., Ellis, R.S., Kneib, J., Le Borgne, J., Pello,
R., Smail, R. \& Sanahuja B., 1997, MNRAS, submitted

\reference{} Ellis, R.S., 1997, ARAA, vol. 35, in press.

%\reference{} Fall, S.M. \& Pei, Y.C., 1993, ApJ, 402, 479

\reference{} Fruchter, A.S. \& Hook, R.N., 1997, 
http://www.stsci.edu/\~{}fruchter/dither/drizzle.html.

\reference{} Gwyn, S.D.J. \& Hartwick, F.D.A., 1996, ApJ 468, L77

\reference{} Gallego, J., Zamorano, J., Aragon-Salamanca, A. \& Rego,
M., 1995, ApJ, 455, L1 

\reference{} Koo, D.C., 1985, AJ, 90, 418

\reference{} Lanzetta, K.M., Yahil, A. \& Fern\'{a}ndez-Soto, A., 1996,
Nature, 381, 759.

\reference{} Lilly, S.J., Le Fevre, O., Hammer, F. \& Crampton, D., 1996, 
ApJ 460, L1

\reference{} Lowenthal, J.D., Koo, D.C., Guzmán, R. Gallego, J.,
Phillips, A.C., Vogt, N.P., Faber, S.M., Illingworth, G.D., \&
Gronwall, C., ApJ, 1997, in press

\reference{} Madau, P., Ferguson, H.C., Dickinson, M.E., Giavalisco,
M., Steidel, C.C. \& Fruchter,A., 1996, MNRAS, 283, 1388

\reference{} Madau, P., 1996, 7th Annual October Astrophysics
    Conference in Maryland, ``Star Formation Near and Far'' in press

\reference{} Mobasher, B., Rowan-Robinson, M., Georgakakis, A. \&
Eaton, N., 1996, MNRAS, 282, L7

\reference{} Pei, Y.C. \& Fall, S.M., 1995, ApJ, 454, 69

\reference{} Sawicki, M.J., Lin, H. \& Yee, H.K.C., 1996, AJ, 117, 1

\reference{} Steidel, C.C., Giavalisco, M., Dickinson, M. \&
Adelberger, K.L., 1996, ApJ, 462, 17

\reference{} Steidel, C.C., Giavalisco, M., Dickinson, M. \& Adelberger, K.L.,
1996, AJ, 112, 352

\reference{} Williams, R.E., Blacker, B., Dickinson, M., Dixon, 
Van Dyke Dixon, W., Ferguson, H.C., Fruchter, A.S., Giavalisco, M.,
Gilliland R.L., Heyer, I., Katsanis, R., Levay, Z., Lucas, R.A., 
McElroy, D.B., Petro, L., Postman, M., Adorf, H.-M., Hook, R.N., 
1996, AJ, 112, 1335.

\end{references}
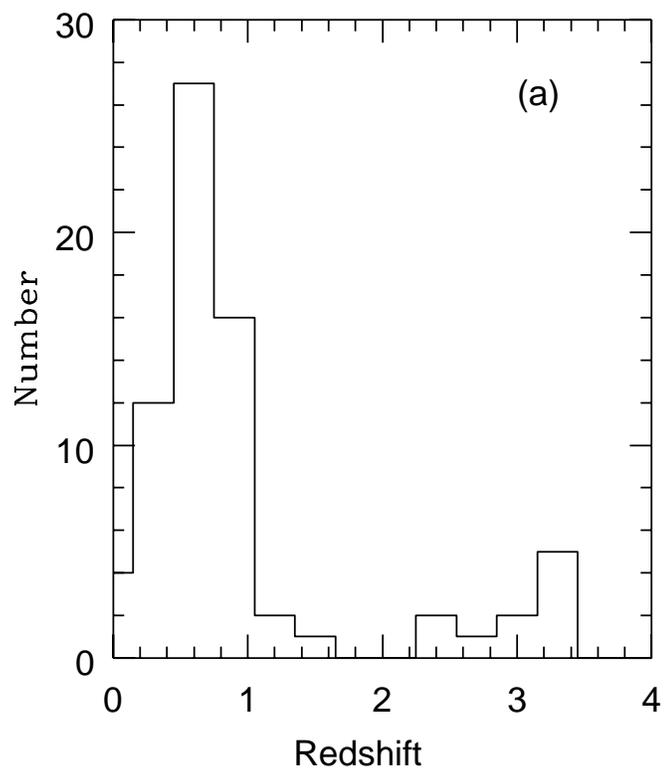
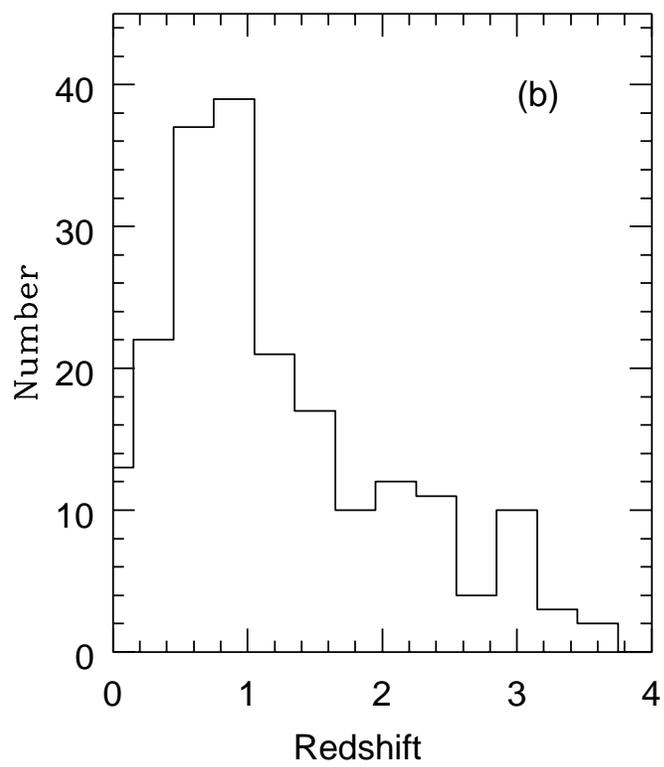
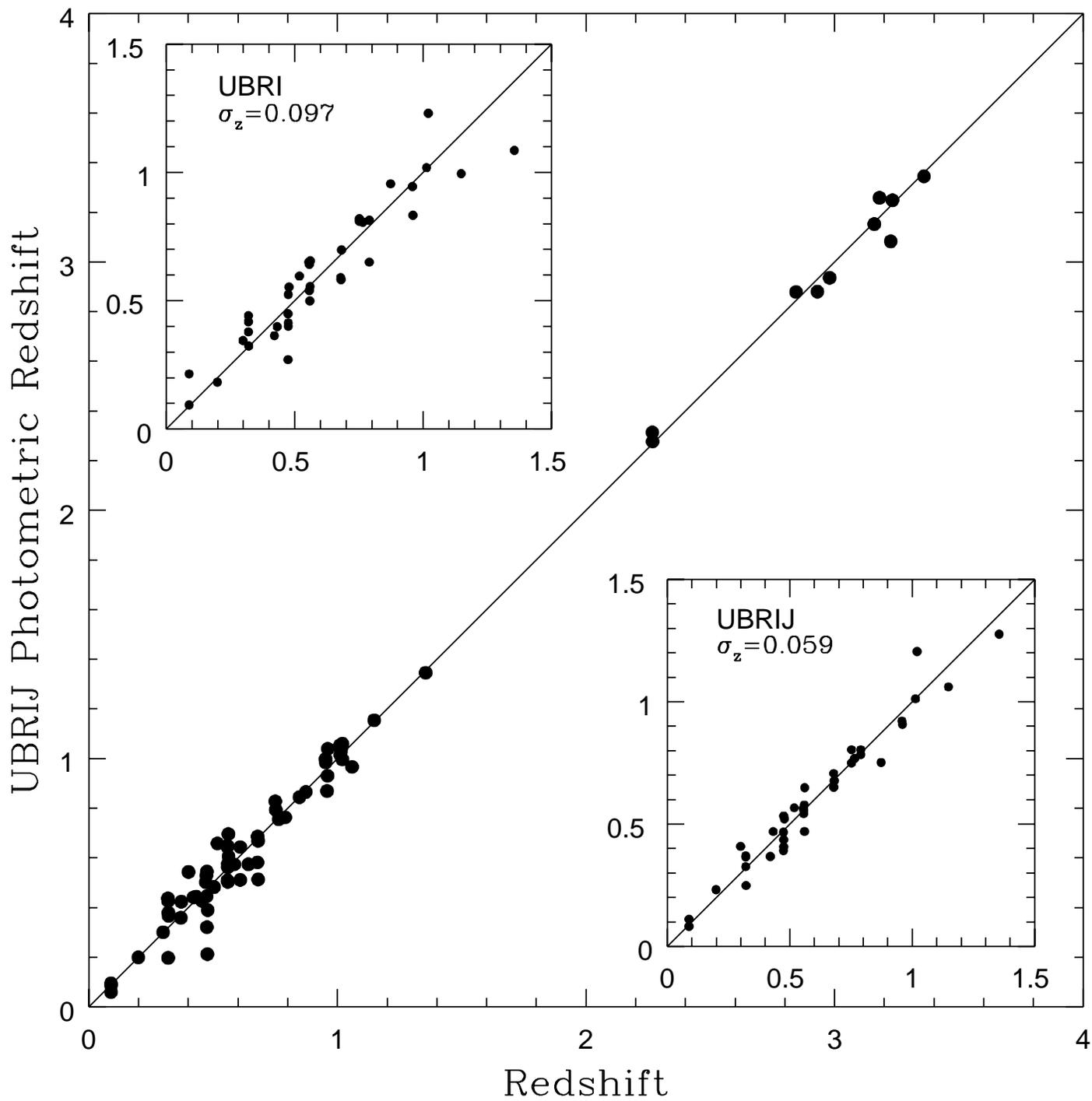
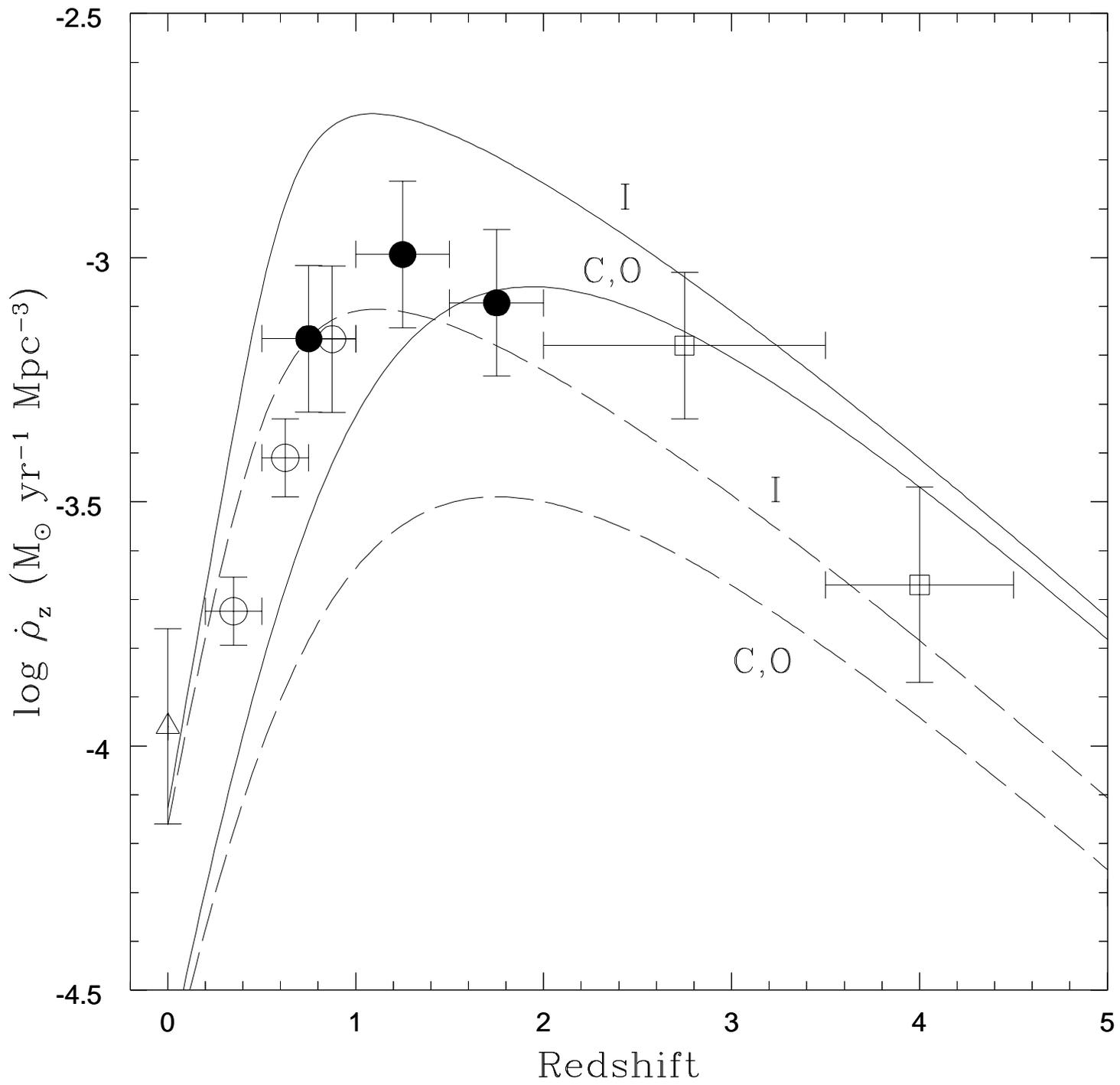
\end{document}